\PassOptionsToPackage{unicode}{hyperref}
\PassOptionsToPackage{hyphens}{url}
\PassOptionsToPackage{dvipsnames,svgnames,x11names}{xcolor}

\documentclass[11pt]{article}

\usepackage{bm}
\usepackage{color}
\def\bSig\mathbf{\Sigma}

\usepackage{amsmath}
\usepackage{amssymb}
\usepackage{amsthm}
\usepackage{subcaption}
\usepackage{float}
\usepackage{booktabs}
\usepackage{multirow}

\usepackage[hidelinks,colorlinks=true,linkcolor=blue,citecolor=blue]{hyperref}

\usepackage{graphicx}
\usepackage{subcaption}
\usepackage{natbib}
\setcitestyle{authoryear,open={(},close={)}}
\usepackage{booktabs}
\usepackage{amsfonts}
\usepackage{amssymb}
\usepackage{dsfont}
\usepackage{eucal}
\usepackage{multirow}
\usepackage[dvipsnames]{xcolor}
\usepackage{caption}
\usepackage{xr-hyper}

\makeatletter
\newcommand*{\addFileDependency}[1]{%
  \typeout{(#1)}%
  \@addtofilelist{#1}%
  \IfFileExists{#1}{}{\typeout{No file #1.}}
}
\makeatother

\DeclareMathOperator*{\argmin}{arg\,min}

\def\bSig\mathbf{\Sigma}

\usepackage{iftex}
\ifPDFTeX
  \usepackage[T1]{fontenc}
  \usepackage[utf8]{inputenc}
  \usepackage{textcomp} % provide euro and other symbols
\else % if luatex or xetex
  \usepackage{unicode-math}
  \defaultfontfeatures{Scale=MatchLowercase}
  \defaultfontfeatures[\rmfamily]{Ligatures=TeX,Scale=1}
\fi
\usepackage{lmodern}
\ifPDFTeX\else  
\fi
\IfFileExists{upquote.sty}{\usepackage{upquote}}{}
\IfFileExists{microtype.sty}{% use microtype if available
  \usepackage[]{microtype}
  \UseMicrotypeSet[protrusion]{basicmath} % disable protrusion for tt fonts
}{}
\makeatletter
\@ifundefined{KOMAClassName}{% if non-KOMA class
  \IfFileExists{parskip.sty}{%
    \usepackage{parskip}
  }{% else
    \setlength{\parindent}{0pt}
    \setlength{\parskip}{6pt plus 2pt minus 1pt}}
}{% if KOMA class
  \KOMAoptions{parskip=half}}
\makeatother
\usepackage{xcolor}
\makeatletter
\ifx\paragraph\undefined\else
  \let\oldparagraph\paragraph
  \renewcommand{\paragraph}{
    \@ifstar
      \xxxParagraphStar
      \xxxParagraphNoStar
  }
  \newcommand{\xxxParagraphStar}[1]{\oldparagraph*{#1}\mbox{}}
  \newcommand{\xxxParagraphNoStar}[1]{\oldparagraph{#1}\mbox{}}
\fi
\ifx\subparagraph\undefined\else
  \let\oldsubparagraph\subparagraph
  \renewcommand{\subparagraph}{
    \@ifstar
      \xxxSubParagraphStar
      \xxxSubParagraphNoStar
  }
  \newcommand{\xxxSubParagraphStar}[1]{\oldsubparagraph*{#1}\mbox{}}
  \newcommand{\xxxSubParagraphNoStar}[1]{\oldsubparagraph{#1}\mbox{}}
\fi
\makeatother

\usepackage{longtable,booktabs,array}
\usepackage{calc} % for calculating minipage widths
\usepackage{etoolbox}
\makeatletter
\patchcmd\longtable{\par}{\if@noskipsec\mbox{}\fi\par}{}{}
\makeatother
\IfFileExists{footnotehyper.sty}{\usepackage{footnotehyper}}{\usepackage{footnote}}
\makesavenoteenv{longtable}
\usepackage{graphicx}
\makeatletter
\def\maxwidth{\ifdim\Gin@nat@width>\linewidth\linewidth\else\Gin@nat@width\fi}
\def\maxheight{\ifdim\Gin@nat@height>\textheight\textheight\else\Gin@nat@height\fi}
\makeatother
\setkeys{Gin}{width=\maxwidth,height=\maxheight,keepaspectratio}
\makeatletter
\def\fps@figure{htbp}
\makeatother

\makeatletter
\@ifpackageloaded{caption}{}{\usepackage{caption}}
\AtBeginDocument{%
\ifdefined\contentsname
  \renewcommand*\contentsname{Table of contents}
\else
  \newcommand\contentsname{Table of contents}
\fi
\ifdefined\listfigurename
  \renewcommand*\listfigurename{List of Figures}
\else
  \newcommand\listfigurename{List of Figures}
\fi
\ifdefined\listtablename
  \renewcommand*\listtablename{List of Tables}
\else
  \newcommand\listtablename{List of Tables}
\fi
\ifdefined\figurename
  \renewcommand*\figurename{Figure}
\else
  \newcommand\figurename{Figure}
\fi
\ifdefined\tablename
  \renewcommand*\tablename{Table}
\else
  \newcommand\tablename{Table}
\fi
}
\@ifpackageloaded{float}{}{\usepackage{float}}
\floatstyle{ruled}
\@ifundefined{c@chapter}{\newfloat{codelisting}{h}{lop}}{\newfloat{codelisting}{h}{lop}[chapter]}
\floatname{codelisting}{Listing}

\makeatother
\makeatletter
\@ifpackageloaded{caption}{}{\usepackage{caption}}
\@ifpackageloaded{subcaption}{}{\usepackage{subcaption}}
\makeatother

\ifLuaTeX
  \usepackage{selnolig}  % disable illegal ligatures
\fi
\usepackage[]{natbib}
\usepackage{bookmark}

\IfFileExists{xurl.sty}{\usepackage{xurl}}{} % add URL line breaks if available
\hypersetup{
  pdftitle={Title},
  pdfauthor={Author 1; Author 2},
  pdfkeywords={3 to 6 keywords, that do not appear in the title},
  colorlinks=true,
  linkcolor={blue},
  filecolor={Maroon},
  citecolor={Blue},
  urlcolor={Blue},
  pdfcreator={LaTeX via pandoc}}

\usepackage{authblk}
\title{\bf  {Bayesian inference under model misspecification${}^{\dagger}$}}
\author[1]{Jack Jewson}
\affil[1]{Department of Econometrics and Business Statistics \\ Monash University, Victoria, Australia}
\affil[ ]{\textit {\textcolor{blue}{jack.jewson@monash.edu}}}%, piotr.zwiernik@upf.edu}}}
\affil[$\dagger$]{Submitted version accepted for publication in Wiley StatsRef: Statistics Reference Online  \url{https://onlinelibrary.wiley.com/doi/10.1002/9781118445112.stat08674}. %\citep{jewson2026bayesian}
}%, 
\date{March 2026}

\begin{document}

%\doparttoc % Tell to minitoc to generate a toc for the parts
%\faketableofcontents % Run a fake tableofcontents command for the partocs

%\part{} % Start the document part
%\parttoc % Insert the document TOC

%\bibliographystyle{natbib}

\def\spacingset#1{\renewcommand{\baselinestretch}%
{#1}\small\normalsize} \spacingset{1}

\setcounter{Maxaffil}{0}
\renewcommand\Affilfont{\itshape\small}

\spacingset{1.42} % DON'T change the spacing!

\maketitle
\begin{abstract}
 The  likelihood input to a Bayesian analysis {almost} never exactly represents how the data were generated, 
calling into question the validity of posterior inferences.
We review a variational interpretation of the Bayesian posterior as an alternative justification for {its} use under model misspecification, and consider the resulting implications on uncertainty quantification and parameter estimation. We then introduce a range of {techniques} that seek to obtain generalised Bayesian inferences 
that account for 
model misspecification.
\end{abstract}


\spacingset{1.45}

\section{Introduction}\label{sec1}

In the Bayesian paradigm (see \textbf{Bayesian Inference}), inferences about parameters $\theta\in\Theta\subseteq\mathbb{R}^p$ from data $y = (y_1, \ldots, y_n) \in \mathcal{Y}^n$ are made using the posterior distribution (see \textbf{Posterior Distribution: Introduction}). The density of this posterior distribution is given by {the} Bayes' rule as
\begin{equation}
    \pi(\theta\mid y) := \frac{\pi(\theta)p(y; \theta)}{\int_{\Theta}\pi(\theta)p(y; \theta)d\theta},\label{Equ:Bayesrule}
\end{equation}
where $\pi(\theta)$ corresponds to the prior density for $\theta$, and $p(y; \theta)$ is the likelihood that data $y$ were observed under parameter $\theta$. The assumptions underlying inference through \eqref{Equ:Bayesrule} {include} that $\pi(\theta)$ exactly encodes the user's beliefs about the plausibility of values of $\theta$ before the data $y$ were {observed}, and that $p(y; \theta)$ exactly correspond to how the data were generated under parameter value $\theta$, or, at least, how the user believes the data to have been generated. 

%\textbf{I like the idea of a unique distribution alot - this is what I mean by exactly elicit, you can elicit things but not get down to one continious density }

These assumptions are almost never satisfied in practice. While it is possible to elicit  broad structural judgements and some minimal probability statements accurately (\textbf{Elicitation}), eliciting the infinitely many probability statements made by $\pi(\theta)$ is never feasible in the time frame of a particular problem (see e.g. \citet{savage2012foundations, dempster1975subjectivist, goldstein1990influence, o2004probability, ruggeri2005robust, berger2012bayesian}). Similarly, Bayesian analyses increasingly involve high-dimensional and/or observational data, and therefore correctly modelling how the data were generated is challenging. As the famous quote goes ``All models are wrong, ...'' \citep{box1976science, box1987empirical}. Such limitations, however, have not prevented a proliferation of Bayesian methods being applied in practice. 
As a result, {such} Bayesian analyses are {conducted} with functioning priors and likelihoods %$\pi_F(\theta)$ and $p_F(y;\theta)$ 
that approximate what would ideally be used. {Such an approach is meaningful only
if the resulting} functional posterior $\pi(\theta\mid y)$ %$\pi_F(\theta\mid x)$ 
is not too dissimilar to the ideal posterior. That is to say, that the posterior inferences are not driven by the simplifying {functional assumptions} that are made to facilitate the analysis. 

Historically, robustness to the prior (see \textbf{Bayesian Robustness}) has received most attention, as the prior represents {an element specific to Bayesian analysis, it has no counterpart in the frequentist or other
paradigms}. For traditional reviews   see  \citet{berger1990robust, berger1994overview, ruggeri2005robust, berger2012bayesian, giacomini2021robust}, with deeper insights provided in later chapters of \citet{insua2012robust}. 
Instead, the focus here is on robustness to misspecification of the likelihood model and recent progress made in both justifying and improving posterior inference under model misspecification. Up to this point, there exists no all encompassing review of methods to deal with model misspecification in a Bayesian analysis. However, partial reviews are provided by \cite{watson2016approximate} who investigate robustness to model misspecification in a decision making framework, and \cite{knoblauch2022generalized} who use a variational approach to {construct posterior inferences} under model and prior misspecification and computational limitations.

%\begin{itemize}
%    \item Bayesian inference requires correct specification of the priro and likleihood
%    \item Much attention has previously been given to the prior, as this is what appears to differentiate a Bayesian and non-Bayes analysis - cite reviews and previous stats refs
%    \item However, although touched on before, there has rcently been an increased interest in likelihood misspecification, botht he consequence and ways to mitigate - we review these here - 
%    \item Absence of a review paper, beyond \cite{watson2016approximate} and \cite{wu2023comparison} and \cite{walker2013bayesian}
%    \item Something about being agnostic to whether the culprit is the data, or the model, the key is that they don't match 
%    \item Make brief distinction about whether objective (how the data was generated) or subjective (how the user believes the data were generated) - but bioth are inesvuitale misspecified
%    \item Want to say in the intro the Bayesian inference under model misspecification can be seen as a valid, coherent methods to learn about the KLD minimisng parameter \cite{berk1966limiting, walker2013bayesian, bissiri2016general}
%    \item End by saying at least asymptotically prior doesn't matter - \textbf{I think maybe end this Section with saying that at least asymptotically it doesn't matter - it seemed to matter but doesn't really Did Jeremias (JMLR) have a reference that said that this didn't matter!} think \cite{wasserman2014bayesian} has something too, but likelihood very much does!
%\end{itemize}

Before proceeding any further we define what we mean by model misspecification. {A model $p(\cdot; \theta)$, $\theta\in\Theta$} is said to be \emph{correctly specified} for data $y$ if there exists $\theta^{\ast}\in \Theta$ such that $y\sim p(\cdot; \theta^{\ast})$. The model is, therefore, \emph{misspecified} when this is not the case i.e. $y\sim G$ with density/mass function $g(\cdot) \neq p(\cdot;\theta)$ for any $\theta\in\Theta$. 
Throughout, we seek to be agnostic about whether it is the data or the model that `causes' the model to be  misspecified. Outliers, for example, are generally {considered} a problem with the data, but could equally result from the model missing some component of the data. %What is important in general, is that there is a disagreement between the data and the model.
{Lastly, we note} that in finite parametric Bayesian models, model misspecification is somewhat more troublesome than prior misspecification, {while the influence} of the prior decreases with sample size  \cite[see e.g.][]{ghosal2000convergence, shen2001rates, walker2004new, wasserman2014bayesian}, the {effect} of model misspecification generally cannot be alleviated by collecting more data.

\section{Bayesian inference under model misspecification}

\citet{bernardo1994bayesian} coined the term $M$-open to capture situations where \eqref{Equ:Bayesrule} is used to update beliefs when the model is known to be misspecified. 
The clearest, {and} most coherent definition, of what it means to conduct a Bayesian analysis with a misspecified model is provided by \cite{bissiri2016general}, who build upon to work of \cite{zellner1988optimal, zhang2006epsilon} and  \cite{walker2013bayesian}, and is further reviewed and expanded in \cite{knoblauch2022generalized}. \cite{bissiri2016general} identify the posterior in \eqref{Equ:Bayesrule} as the solution to the variational optimisation problem 
\begin{align}
    \pi(\theta\mid y) = \argmin_{q(\theta)\in \mathcal{P}} \mathbb{E}_{q(\theta)}\left[-\log p(y; \theta)\right] + \textrm{KLD}\left(q(\theta)||\pi(\theta)\right),\label{Equ:variational_BayesRule}
\end{align}
where $\mathcal{P}$ is the set of all proper probability densities for $\theta$, and $\textrm{KLD}$ is the Kullback-Leibler divergence.
Unlike the definition of $\pi(\theta\mid y)$ in \eqref{Equ:Bayesrule}, \eqref{Equ:variational_BayesRule} no longer requires the model to be correct and instead provides $\pi(\theta\mid y)$ with a decision theoretic foundation. Candidates for posterior beliefs are evaluated based on their closeness to the prior, as measured by $\textrm{KLD}$, and closeness to the data, as measured by the negative log-likelihood or the `self-information' loss \citep{shannon1948mathematical}. %, and the one achieving the optimal trade-off is the posterior. 
The $\textrm{KLD}$ is motivated as resulting in a coherent update - where posterior beliefs are the same if one first updates {using a} part of the data, and then uses this as the prior to update on the rest, as they would be if they had updated {using} both parts of the data in one go \citep{bissiri2012converting}. The negative log-likelihood, also referred to as the log-score, is motivated as being the only local proper scoring rule \citep{bernardo1994bayesian} - where locality means the scoring rule only depends on the model density for what is observed, and propriety ensures the scoring rule is minimised when the model corresponds to how the data were generated \citep{dawid2007geometry, gneiting2007probabilistic}.

As a result, $\pi(\theta\mid y)$ remains `valid' under model misspecification provided one agrees that \eqref{Equ:variational_BayesRule} adequately measures the suitability of potential posterior beliefs. In the following sections we will examine the impact that the components of \eqref{Equ:variational_BayesRule} have on  posterior uncertainty quantification and parameter estimation, and review the {current} literature that modifies \eqref{Equ:variational_BayesRule} for improved posterior inference under model misspecification.

\section{Robust uncertainty quantification}{\label{Sec:UQ}}

{Under the ideal, that is, correctly specified} likelihood model and standard regularity conditions {\citep[e.g.][]{van2000asymptotic}}, the posterior $\pi(\theta\mid y)$ satisfies {the} Bernstein von-Mises (BvM) theorem{, i.e.,} $\pi(\theta\mid y)$ is asymptotically equivalent to a multivariate Gaussian distribution whose mean is the data generating parameter $\theta^{\ast}$ and whose covariance matrix is the inverse Hessian matrix $H(\theta^{\ast})^{-1}$ with $H(\theta) := -\nabla_{\theta}^2 p(y;\theta)$ \citep{marquis1820theorie, von1936wahrscheinlichkeit, bernstein1946theory}. 
As a result, asymptotically the posterior distribution agrees with the sampling distribution of the maximum likelihood estimate{, meaning} that its credibility intervals are also confidence sets. 

When the likelihood model is misspecified, \citet{chernozhukov2003mcmc, kleijn2012bernstein} show that under regularity conditions, the posterior $\pi(\theta\mid y)$ continues to be asymptotically equivalent to a multivariate Gaussian, {but is centred} at the parameter $\theta_{\textrm{KLD}} := \argmin_{\theta\in\Theta} \textrm{KLD}\left(g|| p(\cdot; \theta)\right)$, and {its (asymptotic) covariance matrix is} $H(\theta_{\textrm{KLD}})^{-1}$. This no longer agrees with the asymptotic distribution of the MLE under model misspecification, which although still centered at $\theta_{\textrm{KLD}}$, has the sandwich covariance matrix \citep[e.g.][]{huber1967behavior}. As a result, under model misspecification posterior credibility intervals are no longer confidence sets. {That is to say,} the uncertainty quantification of the Bayesian posterior is not robust to disagreements between the model and the data \citep{royall2003interpreting, muller2013risk}.

%There is no general result that determines whether $\pi(\theta\mid y)$ is more or less concentrated than it would have been if the model was correctly specified, i.e. if the credibility intervals are wider or thinner than the equivalent confidence sets. However,  \cite{holmes2017assigning} noticed that the data being overdispersed relative to the model can lead to over-concentration, with under-dispersed data can result in an under dispersed posterior.

{To rectify this}, Bayes' rule updating \eqref{Equ:Bayesrule} is `generalised' {by introducing} a scalar $w > 0$, often referred to as the learning rate \citep{walker2001bayesian, zhang2006epsilon}, resulting in generalised posterior 
\begin{equation}
    \pi_w(\theta\mid y) := \frac{\pi(\theta)p(y; \theta)^w}{\int_{\Theta}\pi(\theta)p(y; \theta)^wd\theta} = \frac{\pi(\theta)\exp\left\{w\log p(y; \theta)\right\}}{\int_{\Theta}\pi(\theta)\exp\left\{w\log p(y; \theta)\right\}d\theta}.\label{Equ:temperedBayesrule}
\end{equation}
Setting $w < 1$ leads to a posterior more diffuse than \eqref{Equ:Bayesrule} and setting $w > 1$ results in a more concentrated posterior.
Focus has generally been on $w < 1$ to guard again under-coverage and over-confidence. Intuitively, $w < 1$ down-weights the likelihood contribution to the posterior relative to the prior and, therefore, for a fixed prior increases the robustness to disagreements between the likelihood and the data.  

The generalised posterior in \eqref{Equ:temperedBayesrule} is often referred to as a tempered, fractional or power posterior. 
%or Gibbs posterior \citep{zhang2006information}.
Some specific examples include the safe-Bayes paradigm \citep{grunwald2012safe, grunwald2017inconsistency, heide2020safe} for improved empirical risk in linear modelling, coarsened posteriors \citep{miller2018robust} that {replace} conditioning on the exact data in \eqref{Equ:Bayesrule} with conditioning on data arriving in {its} neighbourhood, and \citet{agnoletto2025bayesian} who show that assuming a quasi-likelihood that only specifies the mean and variance function in linear regression results in \eqref{Equ:temperedBayesrule}. 
Methods to set $w$ include \citet{holmes2017assigning} who consider matching experimental information, \citet{lyddon2018generalized} who calibrate to the Bayesian bootstrap, and \citet{syring2019calibrating} and \cite{matsubara2024generalized} who target coverage under repeated sampling.
\citet{wu2023comparison} provide a recent review and comparison of methods. 

Other modifications to \eqref{Equ:Bayesrule} {that seek to make its uncertainty quantification more robust}  include \citet{giummole2019objective} who directly adjust the curvature of $\log p(y;\theta)$ as a function of $\theta$ and \citet{frazier2023calibrated} who replace the log-likelihood in \eqref{Equ:variational_BayesRule} by a quadratic form involving its gradient and an estimation of the gradient's variance. Although subtly different, both ensure that the asymptotic variance of the posterior agrees with the sandwich covariance of the MLE under model misspecification, which {cannot in general be achieved} for all parameters {by using a single learning rate $w$.}

\section{Robust parameter estimation}

%\textbf{When the model is misspecified suddenly the criteria you actually minimise matters}

The BvM mentioned in Section \ref{Sec:UQ} guarantees that under regularity conditions, if the model is correctly specified then \eqref{Equ:Bayesrule} will concentrate around the parameter that generated the data. 
The result of \citet{kleijn2012bernstein}, alongside \citet{berk1966limiting} and \citet{walker2013bayesian}{, has} established that under model misspecification Bayesian inference learns about $\theta_{\textrm{KLD}}$, the parameter of the model class that is {$\textrm{KLD}$-closest} to the data generating distribution.
However, $\theta_{\textrm{KLD}}$ is generally not robust to changes in the tails of the likelihood model or outlying observations in the data (see e.g. \citet{jewson2018principles, knoblauch2022generalized}). This can be seen by noting that minimising the $\textrm{KLD}$ {is equivalent to} minimising the logarithmic score, $-\log p(y;\theta)$, in expectation over the data generating distribution \citep{dawid2007geometry}, and that $-\log p(y;\theta)\rightarrow \infty$ as $p(y;\theta) \rightarrow 0$. This means that very different inferences are obtained by slightly changing the tails of the model or removing an outlier see e.g. \citet{jewson2024stability}.
%Note, that Bayesian inference under model misspecification is learning about the log-score minimising parameter is also clear from \eqref{Equ:variational_BayesRule}.  

%\cite{bernardo1994bayesian} justify this as inference requires learning about tails

%Conversely,
%The lack of robustness  of the log-score or negative log-likelihood minimising parameter is well known in 
The non-Bayesian literature contains many methods to `robustify' (see \textbf{Robust Estimation}) the maximum likelihood estimator to outliers \citep[see e.g.][]{huber1981robust, hampel2011robust, huber2011robust, basu2011statistical, dawid2016minimum}. 
%%Such methods replace the log-likelihood with an alterative criteria and are generally called $M$-estimation methods. 
%Examples include $M$-estimation methods such as those that replace the squared error loss in linear regression with Huber's \citep{huber1964robust} or Tukey's loss \citet{beaton1974fitting}, using proper scoring rules \citep{dawid2016minimum} different from the logarithmic score, or minimising an alternative divergence between the model and the data generating distribution to the $\textrm{KLD}$ \citep{basu2011statistical}. 
However, only a small subset of these methods can be linked to integrable probability densities that could have generated the data \citep{jewson2022general}, and so Bayesian analogues were prohibited by the {probabilistic foundations} of \eqref{Equ:Bayesrule}.
Instead, traditional Bayesian solutions favour heavy tailed models. For example, replacing a Gaussian distribution with a Student's-$t$ is recommended in  \citet{o1979outlier, west1984outlier, berger1994overview}, while \cite{box1962further, shyamalkumar2000likelihood} consider a class of models with flexible heavy tailed behaviour. More recently, \cite{gagnon2023theoretical} analysed linear regression with Student's-$t$ errors and found cases where its tails were {not sufficiently heavy to ignore} outliers. This motivated the `log-Pareto-tailed normal' distribution as an even heavier tailed alternative \citep{gagnon2020new}.

However, the representation provided by \eqref{Equ:variational_BayesRule} no longer requires $\theta$ to index a probability model. \cite{zhang2006information} and \cite{bissiri2016general} took advantage of this and proposed a generalised Bayesian updating of beliefs via  
\begin{align}
    \pi_{\ell}(\theta\mid y) :&= \argmin_{q(\theta)\in \mathcal{P}} \mathbb{E}_{q(\theta)}\left[\ell(y,  \theta)\right] + \textrm{KLD}\left(q(\theta)||\pi(\theta)\right)
    %\label{Equ:variational_GeneralisedBayes}\\
    %&
    = \frac{\pi(\theta)\exp\left\{-\ell(y,  \theta)\right\}}{\int_{\Theta}\pi(\theta)\exp\left\{-\ell(y,  \theta)\right\}d\theta},\label{Equ:GeneralisedBayes}
\end{align}
where $\ell(y,  \theta)$ is a loss function that evaluates the suitability of parameter $\theta$ for observed data $y$. The posterior in \eqref{Equ:GeneralisedBayes} is often referred to as a Gibbs posterior \cite[e.g.][]{zhang2006information, jiang2008gibbs}. 
%\textbf{Zhang 2006a (power-postereior) Zhange 2006b - show that Gibbs posterior minimises same criteria as Bissiri!! Zhang seems to have been the first person to proopse this!}

%The formulation in \eqref{Equ:GeneralisedBayes} has subsequently allowed for Bayesian implementations of the traditional robust estimation procedures mentioned above. 
\citet{jiang2008gibbs} took $\ell(y,  \theta)$ to be the misclassification error in logistic regression and proved that this led to improved predictive performance under misspecification, \citet{loaiza2021focused} recently extended this beyond classification. Around the same time \citet{greco2008robust} proposed an update similar to \eqref{Equ:GeneralisedBayes} using $M$-estimators, such as those that replace the squared error loss in linear regression with Huber's \citep{huber1964robust} or Tukey's loss \citet{beaton1974fitting}. Bayesian $M$-estimation has been further formalised in \cite{marusic2025theoretical}, see also \cite{lewis2021bayesian} who incorporate $M$-estimates via {a framework that allows for conditioning on statistics that are not sufficient} and \cite{jewson2022general} who propose Bayesian methods to select the $M$-estimation hyperparameters. 
{\citet{syring2017gibbs} used \eqref{Equ:GeneralisedBayes} to circumvent the need for a model when estimating the minimum clinically important difference of a treatment effect.}

%When inference for the data-generating process is the goal \textbf{IMPROVE}, 
A particularly appealing class of loss functions {results} from trying to minimise a divergence, $D(\cdot || \cdot)$ {between} the data generating process, $g$, {and the model}, $p(\cdot; \theta)$, that is different from the $\textrm{KLD}$ associated with the logarithmic score in \eqref{Equ:variational_BayesRule} \citep{jewson2018principles}. 
%So doing allows one to continue to learn the same model parameters but obtain different robustness properties to the log-score. 
Early examples include using the Hellinger \citep{hooker2014bayesian} or total-variation distance \citep{jewson2018principles}, but these required a non-parametric density estimate to evaluate the loss function, which prevents them scaling beyond low-dimensional problems. 
More convenient are divergences whose minimisation can be replaced by minimising a proper scoring rule (see \textbf{Scoring Rules}) (as $\textrm{KLD}$ was replaced by log-score) \citep{giummole2019objective}. Examples include the $\beta$-divergence which is also known as the density power divergence and is associated with the Tsallis score \citep{ghosh2016robust, jewson2024stability}, Fisher's divergence which is associated with the {Hyv\"arinen} score \citep{jewson2022general, altamirano2023robust} and the $\gamma$-divergence \citep{hung2018robust, knoblauch2022generalized}.
Lastly, there are discrepancy measures such as the maximum mean discrepancy \citep{cherief2020mmd} and  kernel-Stein discrepancies  \citep{matsubara2022robust} which omit loss functions that compare simulations from the model with data $y$ and thus allow for inference on models with intractable densities as well as providing robustness to model misspecification.

%\textbf{Divergence and propoer score aim to learn the same data generating process if correctly specified, M-estimation methods the same true parameter}

Many of the loss functions mentioned above {would} still learn about the data generating parameters if the model were not misspecified, with some loss of statistical efficiency.
For example, it is a specific property of divergences, discrepancies and proper-scoring rules that they {have unique minima} when the model and the data generating process {coincide}. 
%
%Instead, it is when the model is misspecified that these loss function, although based on the same data, and often the same model, learn about different parameters with different robustness properties to $\theta_{\textrm{KLD}}$.
{However, these loss functions, even when based on the same data, and often the same model,
learn about different parameters with different robustness properties to $\theta_{\textrm{KLD}}$ when the model is misspecified.}

%Examples of this include Bayesian empirical risk minimisation \citep{zhang2006information, jiang2008gibbs}, Bayesian methods that learning about parameters that minimiser alternative divergences to the KLD \citep{jewson2018principles, giummole2019objective, knoblauch2022generalized}, with specific examples include the Hellinger divergence \citep{hooker2014bayesian}, $\beta$-divergence (which is also referred to as the density power divergence and is related to the Tsallis proper scoring rule \citep{dawid2016minimum}) \citep{ghosh2016robust, giummole2019objective}, $\gamma$-divergence \citep{hung2018robust}, Fisher divergence which is related to the Hyvarinen score (also referred to as score matching) \citep{giummole2019objective, jewson2022general, matsubara2022robust}, kernel Stein \citep{matsubara2022robust} and maximum mean discrepancies \citep{cherief2020mmd}, as well as more general M-estimators \citep{greco2008robust, marusic2025theoretical, lewis2021bayesian}.

{Note} that while \eqref{Equ:variational_BayesRule} allows flexibility in choosing a loss function, it provides no guarantees that the uncertainty quantification provided by \eqref{Equ:GeneralisedBayes} is in any way meaningful. In particular, replacing the log-likelihood with a loss function often results in a change of scale. Probability densities must integrate to 1, but there is no such {constraint} for a loss function. {Multiplying the loss by a constant $C > 0$, $C \neq 1$, } does not change the parameter the loss learns about, but does change %the relative weight of the prior and data terms in the variational optimisation which changes 
the uncertainty quantification of \eqref{Equ:GeneralisedBayes}.
As a result, the scale of the loss needs `calibrating'. 

A popular strategy places a learning-rate parameter, $w > 0$, in front of the loss, {similarly to $w$ in \eqref{Equ:temperedBayesrule}}. 
Under regularity conditions \cite{chernozhukov2003mcmc} {and \cite{miller2021asymptotic}} provide a Bernstein-von-Mises theorem for \eqref{Equ:GeneralisedBayes}, where the asymptotic variance is the inverse of the second derivative of the loss function. This does not agree with the repeated sampling asymptotic distribution of the loss minimiser \citep[e.g.][]{van2000asymptotic}. 
Methods have, therefore, generally {focused} on setting $w$ to better match the repeated sampling asymptotics \citep{lyddon2018generalized, syring2019calibrating, matsubara2024generalized} or modifying the loss function to elicit the same asymptotic distribution \citep{giummole2019objective, frazier2023calibrated}.
Note {that} this literature generally does not make a  distinction between using the log-score with a misspecified model, or a general loss function.

%\textbf{Gibbs posteriors - Jiang and Tanner!} Talk about them doing Bayesian empriical risk minimisation motvated by model misspecification! first application?

%mention general results about epsilon contamination, more general result for the beta-divergence in my aper 

\section{Sensitivity Analysis}

%The previous two sections have introduced inherently robust Bayesian procedures that generalise the standard Bayesian posteriro \eqref{Equ:Bayesrule}.
For completeness, we review the traditional approaches to analyse the robustness of a specific model specification, generally referred to as a sensitivity analysis. {As pointed out in \cite{berger2012bayesian}}, while such methods are better associated with prior robustness, they can be adapted to {explore perturbations of} the model. The most extensive example of this was provided by \cite{lavine1991sensitivity}, however such formal analyses are rare.

The general strategy is to investigate the extent to which the conclusions of an analysis, for example a  posterior estimate or credibility interval, would change if a {slightly different functioning model had been applied}. 
If they vary a lot {then the likelihood
specification has to be reviewed}.
Briefly, approaches here can be classified as unsystematic, where some alternative models that immediately come to mind are implemented, but such analyses remain ad-hoc. More formally, one could construct a parametric class around the model capturing perturbations of it (e.g. $\epsilon$-contaminations \citep{huber1964robust}). {They assess} so called global robustness. 
A particular challenge with global methods is that different models often do not have equivalent parameters, and, therefore, comparing posteriors becomes challenging. {An approach} investigates posterior predictive stability \citep{jewson2024stability}. 
%One solution, however, is to investigate the stability of posterior predictive rather than posterior  \citep{jewson2024stability}. 
Alternatively, one could investigate the impact of infinitesimal departures from the model, so called local robustness.  One output of local analyses are posterior influence functions which, for example, are considered in \cite{greco2008robust, ghosh2016robust} and \cite{matsubara2022robust}.

%\begin{itemize}
%    \item For completeness we review traditional sensitivity first
%    \item Can use all prior sensitivity methods for likelihood \cite[e.g.][]{berger2012bayesian}
%    \item \cite{lavine1991sensitivity} - maybe also huber contaminations!
%    \item problems with the parameter changing 
%    \item Mention posterior influence functions here 
%\end{itemize}

%\textbf{Would be great to mention influence functions here} \cite{greco2008robust} discusses posteriro influence function, so do \cite{matsubara2022robust} I think and \cite{ghosh2016robust}

\section{Discussion}

%We reviewed a justification from variational optimisation for the use of Bayes' rule posterior inference under model misspecification, and introduced a variety of methods that seek to generalise and improve its uncertainty quantification and parameter estimation to accommodate the inevitability of model misspecification. % acknowledging the inevitability of model misspecification. 
%
We have reviewed the variational foundations of Bayesian inference in the presence of model misspecification and introduced a range of methods designed to robustify parameter estimation and uncertainty quantification.
At the time of writing this {is} a new and fast moving field. 
Related literature that we have not discussed includes predictive Bayesian approaches %that choose not to specify a prior or likelihood altogether 
\citep{fong2023martingale, holmes_statistical_2023, fortini2023prediction} - see \cite{fortini2025exchangeability} for a review; predictive extensions to the optimisation problems in \eqref{Equ:variational_BayesRule} and \eqref{Equ:GeneralisedBayes} \citep{lai2024predictive, shen2025prediction, mclatchie2025predictively}; $e$-posteriors based on $e$-values \citep{grunwald2023posterior}; and connections to PAC-Bayes \citep{guedj2019primer, alquier2024user, viallard2024leveraging}.

Clearly, a better specified model would remove many of the issues discussed here, and there exists a multitude of flexible modelling techniques which include Bayesian non-parametrics and machine learning. 
However, such a model should not come at the expense of the structure a parametric model can impose and the interpretability it can provide, or be over-burdensome computationally.
It should further be acknowledged {that misspecification
cannot be ruled out even in some very complex models}.
%that even the most complicated models are likely still misspecified in some way relative to how the data were generated. 

%We conclude with a final, important observation. While, for the reasons discussed here it is important that Bayesian analyses are robust, the goal for robustness must always be balanced with the, sometimes, competing goal of learning form data. Estimating a parameter to always be 7, independently of the data or prior is completely robust to their specification, but is clearly not very useful. Perhaps more realistically, setting the posterior to the prior is a perfectly valid was to be robust to the likelihood model, but again prevents learning form data. 

\section*{Acknowledgements}

The author would like to thank Dr. Nicolas Longford for editorial feedback.

\bibliography{bib.bib}

\end{document}